# Effect of Coulomb interaction on the two-dimensional electronic structure of the van der Waals ferromagnet $Cr_2Ge_2Te_6$

M. Suzuki,[1*] B. Gao,[2] K. Koshiishi,[1] S. Nakata,[1] K. Hagiwara,[1] C. Lin,[1] Y. X. Wan,[1] H. Kumigashira,[3] K. Ono,[3] Sungmo Kang,[4] Seungjin Kang,[4] J. Yu,[4] M. Kobayashi,[5,6] S-W. Cheong,[2] and A. Fujimori[1]

[1]*Department of Physics, The University of Tokyo, Bunkyo-ku, Tokyo 113-0033, Japan*

[2]*Rutgers Center for Emergent Materials and Department of Physics and Astronomy, Rutgers University, Piscataway, New Jersey 08854, USA*

[3]*Photon Factory, High Energy Accelerator Research Organization (KEK), Tsukuba, Ibaraki 305-0801, Japan*

[4] *Center for Theoretical Physics, Department of Physics and Astronomy, Seoul National University, Seoul 08826, Republic of Korea*

[5]*Center for Spintronics Research Network, The University of Tokyo, Bunkyo-ku, Tokyo 113-0033, Japan*

[6]*Department of Electrical Engineering and Information Systems, The University of Tokyo, 7-3-1 Hongo, Bunkyo-ku, Tokyo 113-8656, Japan*

In order to investigate the electronic properties of the semiconducting van der Waals ferromagnet $Cr_2Ge_2Te_6$ (CGT), where ferromagnetic layers are bonded through van der Waals forces, we have performed angle-resolved photoemission spectroscopy (ARPES) measurements and density-functional-theory (DFT+$U$) calculations. The valence-band maximum at the $\Gamma$ point is located ~ 0.2 eV below the Fermi level, consistent with the semiconducting property of CGT. Comparison of the experimental density of states with the DFT calculation has suggested that Coulomb interaction between the Cr 3$d$ electrons $U_{eff}$ ~ 1.1 eV. The DFT+$U$ calculation indicates that magnetic coupling between Cr atoms within the layer is ferromagnetic if Coulomb $U_{eff}$ is smaller than 3.0 eV and that the inter-layer coupling is ferromagnetic below $U_{eff}$ ~ 1.0 eV. We therefore conclude that, for $U_{eff}$ deduced by the experiment, the intra-layer Cr-Cr coupling is ferromagnetic and the inter-layer coupling is near the boundary between ferromagnetic and antiferromagnetic, which means experimentally deduced $U_{eff}$ is consistent with theoretical ferromagnetic condition.

# I. INTRODUCTION

Since the discovery of graphene, there has been tremendous interest in the development of new two-dimensional (2D) materials and their functionalities [1,2]. The materials have layered crystal structures and the layers are bonded to each other through van der Waals (vdW) forces. Due to the reduction of dimensionality in the electronic states, 2D materials show fascinating quantum properties which would lead to various potential applications for new devices [3,4]. In addition to graphene and transition-metal dichalcogenides, vdW ferromagnets have recently attracted much attention as candidate materials for new types of spintronic devices.

$Cr_2Ge_2Te_6$ (CGT) is a representative vdW ferromagnet and, along with other vdW ferromagnets, *e.g.*, $CrI_3$ and $Cr_2Si_2Te_6$, its electronic and magnetic properties have been studied theoretically and experimentally [5 - 12]. The Curie temperature of CGT has been reported to be of 61 K [5 - 7] and the magnetic moment of 2.2 - 3.0 $\mu_B$/Cr atom with the easy magnetization axis along the *c*-axis (out-of-plane direction) [5,8]. As for the electric properties, resistivity measurements on CGT have shown semiconducting behaviors and its resistivity prominently increases at low temperatures [5,8]. In their bulk forms, CGT shows physical properties similar to $CrI_3$. The magnetism of CGT nanosheets shows soft behaviors and is well described by the 2D Heisenberg model [9]. In contrast to the Heisenberg description of the CGT nanosheet, ferromagnetism in monolayer of $CrI_3$ is well described by the 2D Ising model [10]. The difference in the magnetic behaviors between $CrI_3$ and CGT could may stem the different strengths of inter-layer magnetic coupling and the different electronic structures of Cr atoms.

To unveil the origin of the ferromagnetism in 2D materials, understanding the physical properties from the electronic structure points of view is important. Indeed, both theoretical and experimental studies on 2D ferromagnetic materials have been made in recent years [3-15]. Band-structure calculations on CGT have shown a semiconducting property, which is consistent with experiment. For both fundamental understanding and potential applications of 2D vdW ferromagnets, the knowledge of the electronic structure of CGT related to the ferromagnetism is indispensable. In the present study, we have investigated the electronic properties of CGT single crystals using angle-resolved photoelectron spectroscopy (ARPES) and compared the experimental result with DFT+$U$ calculation. The results suggest that the Coulomb interaction plays an important role in the magnetic properties of CGT.

## II. METHODS

Single crystals of CGT were synthesized using the flux method. High purity Cr powders, Ge powders, and Te powders were mixed in a molar ratio of 2:6:36; the extra Ge and Te were used as a flux. The mixture was loaded in an alumina crucible and sealed in an evacuated quartz tube. A ball of alumina fiber was placed on top of the crucible. The quartz tube was placed in a furnace, heated to 700 °C, held for 5 hours, then slowly cooled down to 480 °C over a period of 3 days. This was followed by centrifugation to remove the flux.

ARPES measurements were performed at beam line BL-28A of Photon Factory, KEK. The incident light was circularly polarized and the energy range used was 34-80 eV. A SCIENTA SES 2002 electron analyzer (Scienta-Omicron Ltd.) was used. Measurements were performed under the vacuum of $2.0 \times 10^{-8}$ Pa and the sample was cleaved in the analyzer chamber prior to the measurements to obtain fresh surfaces. The sample temperature during the measurements was set to 150 K, above the transition temperature, in order to avoid charging effects. Binding energies were calibrated by measuring the Fermi level ($E_F$) of an Au foil that is in electrical contact with the sample. The energy resolution was set to ~ 50 meV.

First-principles band-structure calculations were conducted based on DFT using the full-potential augmented-plane-wave method implemented in the WIEN2K and OpenMX packages. By the WIEN2k package, the density of states (DOS) and the band structures of the spin-polarized state were calculated. The lattice constants of $a = b = 6.8275$ Å and $c = 20.5619$ Å were used. Generalized-gradient-approximations (GGA) calculation and GGA + $U$ calculation, which leads to orbital dependent potentials for the Cr 3$d$ orbitals, were conducted for bulk materials. Furthermore, using the OpenMX package, local-density-approximation (LDA) and LDA + $U$ calculations were performed for bulk and single-layer CGT in order to investigate the relationships between inter- and intra-layer magnetic coupling and Coulomb interaction. Calculations by WIEN2k and OpenMX for the same Coulomb $U$ give almost the same results.

## III. RESLUTS AND DISCUSSION

Figure 1(a) shows a constant energy surface mapping at $E$ = - 0.35 eV, where the intensity has been integrated within the energy window of ± 0.25 eV. There is a circular constant energy surface around the Γ point. Figure 1(b) shows ARPES spectra along the cut shown by a red line in Fig. 1(a) around the Γ point. In order to closely look at the dispersion around the valence-band maximum (VBM), ARPES spectra enlarged around the Γ point and the corresponding energy distribution curves (EDC) are shown in Figs.1 (c) and (d), respectively. In Figs. 1 (e) and (f), we compare band dispersions around the VBM taken using three photon energies 34, 56, and 80 eV corresponding to $k_z \sim 17.46$, 23.07, and 27.94 π/c. Here, $k_z$'s are estimated using the inner potential of 10 eV. The almost identical band dispersions for the three photon energies indicate that dispersions along the $k_z$ direction are negligibly small in this compound and that the electronic structure of CGT is highly 2D reflecting its layered crystal structure.

The result demonstrates that there is a hole-like band dispersion from $E$ = - 1 eV to $E_F$ and some dispersive bands around $E$ = - (2 - 5) eV. A clear gap is observed below $E_F$, consistent with the semiconducting property [5,8]. The VBM is located at ~ 0.23 eV below $E_F$. Because the band gap of 0.20 - 0.24 eV has been deduced from the resistivity measurements in the intrinsic conduction region (T > ~ 100 K) [5,8], the $E_F$ position in the ARPES spectra is just below the conduction band minimum, which contradicts with the fact that CGT should be an intrinsic semiconductor around 150 K, where the ARPES measurements were made. Reported electronic properties show that CGT is an intrinsic semiconductor unless vacancies do not exist at the Ge sites [5,8]. Probably there would be a band bending which lowers both the valence and conduction bands at the surface and only the bended band has been observed due to a short probing depth of ARPES.

In order to investigate the electronic structure in more detail, we have conducted spin-polarized DFT calculation for the ferromagnetic phase of this compound and compared the calculated results with the ARPES spectra. While the present ARPES measurements have been performed above the Curie temperature $T_C$, we consider that the exchange splitting of the ferromagnetic state persists above $T_C$ and that the electronic structure in the paramagnetic phase is more similar to that calculated for the ferromagnetic state rather than that calculated for the paramagnetic state [16-18]. Figure 2 shows the spectra of Fig. 1 (b) integrated with respect to $k$ along the cut shown in Fig. 1(a) compared with the calculated $k$-integrated spectrum. Here, the photoionization cross-sections of the constituent atoms are taken into account. The cross-section of the Cr 3$d$ orbital is larger than those of the other orbitals in the valence band for the photon energy of 80 eV. A prominent peak appears in the experimental spectrum around ~ 1.9 eV, which is significantly deeper than the peak position of ~ 1.3 eV in the calculated one without Coulomb interaction. In order to resolve this discrepancy, we have taken into account Coulomb interaction $U$ between Cr 3*d* electrons in the DFT calculation. We define the effective Coulomb interaction as $U_{eff} = U - J$, where $U$ is the atomic Coulomb integral and $J$ is the exchange parameter. Fig. 2 shows that calculation with $U_{eff}$ = 1.1 eV well reproduces the experiment.

To examine the effects of Coulomb interaction on the valence-band dispersions, the obtained band dispersions are compared with the calculation. Figures 3(a) and (b) show ARPES spectra around the VBM compared with the calculated band dispersions. The band dispersions from the calculation well reproduce the ARPES spectra except for the heaviest bands near the VBM. This is probably because the cross-sections of the bands are not large enough to be observed for the photon energies employed in the present work. It should be noted that the calculated band dispersions are nearly independent of the Coulomb interaction $U_{eff}$ between Cr 3*d* electrons. According to the calculation, the Te 5*p* states are dominant around the VBM. Furthermore, the width of the entire valence band is seemingly not renormalized even in the presence of the Coulomb interaction, as shown in Figs. 3. These results indicate that the Coulomb interaction affects the Cr 3*d* states but hardly the Te 5*p*- and the Ge 3*p*-derived valence bands.

In a previous DFT calculation on $Cr_2O_3$, it has been reported that the value of $U_{eff}$ for Cr 3*d* electrons is about 3.5 eV [19]. The present calculation on CGT on the other hand, indicates that the magnetic coupling within the layer is ferromagnetic for $U_{eff}$ < 3.0 eV, and becomes antiferromagnetic for $U_{eff}$ > 3.0 eV [20]. On the other hand, inter-layer coupling is ferromagnetic for $U_{eff}$ < 1.0 eV and antiferromagnetic for $U_{eff}$ > 1 eV [20]. $U_{eff}$ = 1.1 eV estimated in the present experiment well explains the ferromagnetic intra-layer coupling while the inter-layer magnetic coupling is expected to be near the boundary between ferromagnetic and antiferromagnetic. Thus, the present results indicate the importance of Coulomb interaction on the magnetic and electronic structures of CGT.

In the present study, we have confirmed that the valence−band structure of CGT is 2D and the Coulomb interaction affects the Cr 3*d* states but not the Te 5*p* and Ge 3*p* states. In $CrI_3$, it has been found that ferromagnetism persists even in a monolayer thickness but that the inter-layer coupling becomes antiferromagnetic in bilayers, i.e., the inter-layer interaction becomes ferromagnetic or antiferromagnetic depending on the layer number of $CrI_3$ [10]. CGT maintains the soft ferromagnetic properties even in ultrathin films of bilayers [9] and sign change of the inter-layer magnetic coupling has not been observed. This difference probably comes from different environment around the Cr atom. Comparison of the experimental results with the calculation on CGT suggests that the presence of Coulomb interaction makes the energy position of the Cr 3*d* states deeper, *i.e.*, Coulomb interaction localizes the Cr 3*d* states and plays a key role in the magnetic properties of CGT. To understand highly 2D vdW ferromagnets from the electronic structure point of view, further studies are desirable. For example, comparison of the electronic structure, especially from the viewpoint of Coulomb interaction, between CGT and $CrI_3$ will shed light on the difference of magnetic properties between CGT and $CrI_3$ in the 2D limit.

## IV. CONCLUSION

We have performed ARPES measurements and DFT calculation on the vdW ferromagnetic semiconductor CGT in order to investigate its electronic structure. From the ARPES spectra, it is found that the VBM does not reach the $E_F$, consistent with the semiconducting behavior of this compound. The DFT+$U$ calculation well explains the ARPES results. By comparing the experimental results with calculation, it is suggested that the estimated value of Coulomb interaction $U_{eff}$ favors intra-layer ferromagnetism and is near the boundary between inter-layer ferromagnetic and antiferromagnetic coupling. This suggests the importance of Coulomb interaction between the Cr 3*d* electrons for understanding both the electronic structure and magnetic properties of CGT. It is also found that the top of the valence band is mainly composed of non-magnetic orbitals, Te 5*p*, and is not strongly affected by the Coulomb interaction between Cr 3*d* electrons. Our study has revealed that Coulomb interaction plays an important role not only in its electronic structure but also in its magnetic properties.

**Acknowledgments**

This work was supported by a Grant-in-Aid for Scientific Research from JSPS (grant No. 15H02109) from MEXT and by Center for Spintronics Research Network (CSRN), the University of Tokyo, under Spintronics Research Network of Japan (Spin-RNJ). AF is an adjunct member of CSRN. The synchrotron experiment was done under the approval of the Program Advisory Committee (Proposal Nos. 2015S2-003, 2016G096, and 2018G049). The work at Rutgers University was supported by the NSF under Grant No. DMR-1629059.

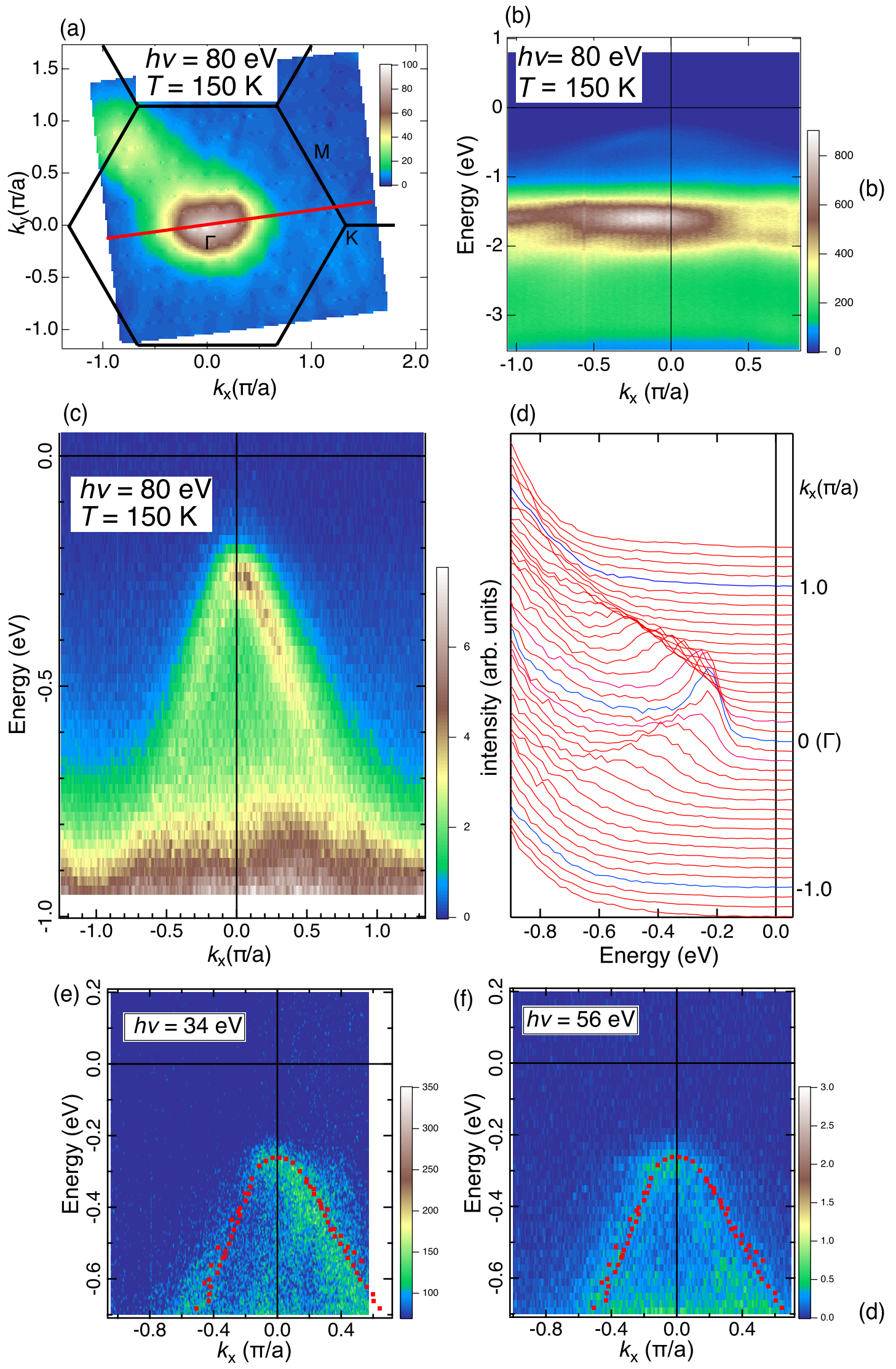

**FIG. 1:** ARPES spectra of $Cr_2Ge_2Te_6$. (a) Constant energy surface mapping at $E$ = - 0.35 eV. The intensity has been integrated within the energy window of $\Delta E$ = +/-0.25 eV. The used photon energy was 80 eV. The hexagon indicates the Brillouin zone. (b) $E$-$k$ image along the cut though the Γ point shown by a red line in (a). (c) Enlarged $E$-$k$ image around the valence-band maximum (VBM) at the Γ point. The hole-like band is clearly observed and the VBM is about 0.23 eV below the Fermi level ($E_F$), reflecting the semiconducting property of this compound. (d) Energy distribution curves of the data shown in panel (c). (e)(f) ARPES spectra taken with the photon energies of 34 eV and 56 eV. Red dots indicate the peak positions of the top of the valence band in the ARPES spectra taken with the photon energy of 80 eV.

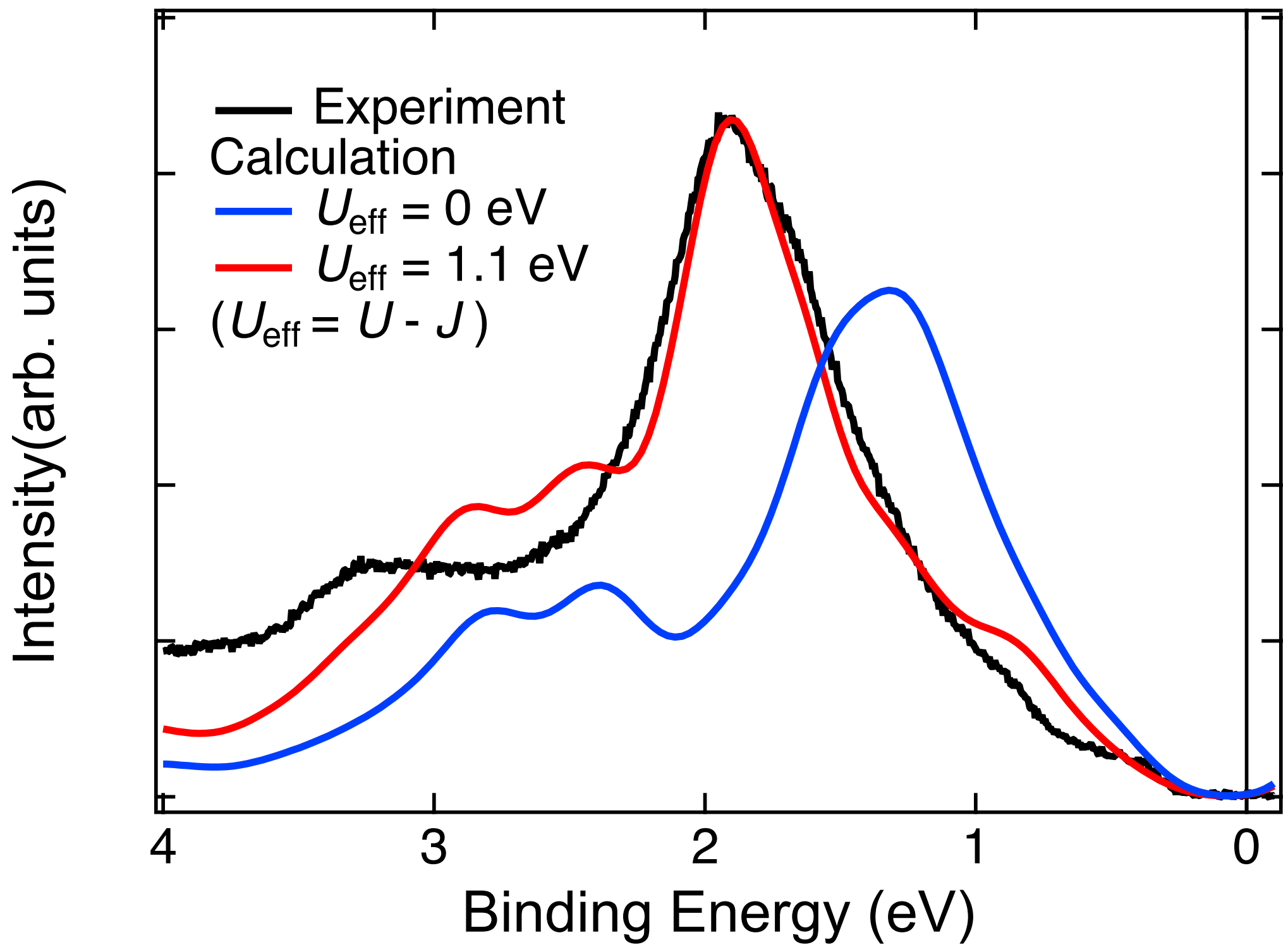

**FIG. 2:** $k$-integrated photoemission spectrum (black curve) taken with the photon energy of 80 eV compared with calculated spectra. The $k$-integration has been made along the cut shown in Fig. 1(a). Calculated result without Coulomb interaction and that with effective Coulomb interaction $U_{\text{eff}} = 1.1$ eV are compared with experiment. The calculation takes into account the cross sections of atomic orbitals for the photon energy of 80 eV.

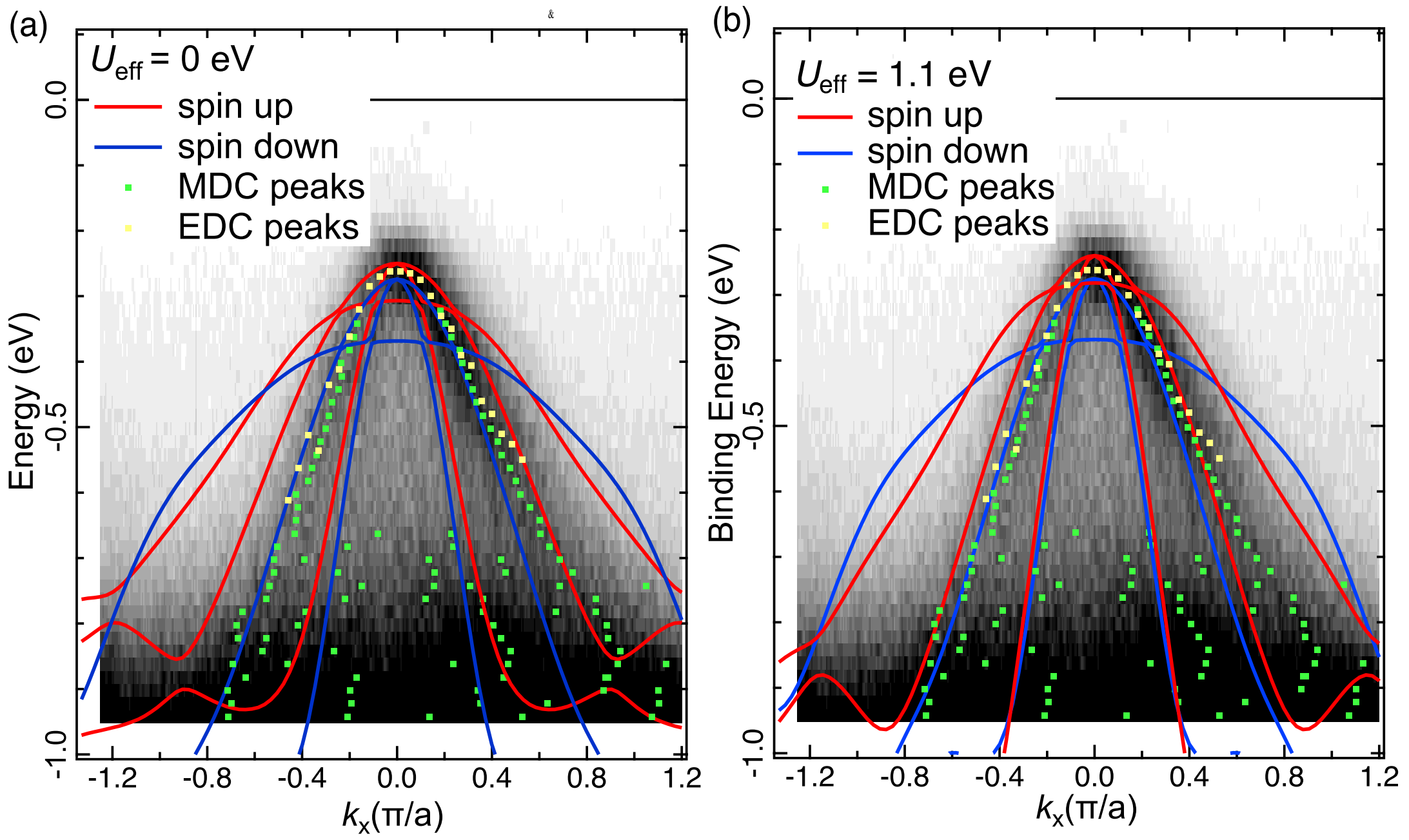

**FIG. 3:** ARPES spectra around the VBM at the Γ point compared with the band-structure calculation (a) without and (b) with $U_{\rm eff}$ =1.1 eV. Curves indicate spin-up and spin-down bands. Yellow and green dots are the peak positions obtained from the second derivatives of the EDCs and MDCs, respectively.